\documentclass[12pt]{iopart}
\usepackage{subfigure}
\usepackage{iopams}
\usepackage{graphicx}

\newcommand{\hc}{\hat{h}_{source}}

\begin{document}

\title{LISA Parameter Estimation Using Numerical Merger Waveforms}

\author{J.I. Thorpe$^{1}$, S.T. McWilliams$^{1}$, B.J. Kelly$^{1}$, R.P. Fahey$^{1}$, K. Arnaud$^{1,2}$,
and J.G. Baker$^{1}$}
\address{$^{1}$ NASA Goddard Space Flight Center, 8800 Greenbelt Rd., Greenbelt, MD 20771, USA}
\address{$^{2}$ CRESST and Department of Astronomy, University of Maryland, College Park, MD 20742, USA}
\ead{James.I.Thorpe@nasa.gov}

\begin{abstract}
Recent advances in numerical relativity provide a detailed description
of the waveforms of coalescing massive black hole binaries (MBHBs),
expected to be the strongest detectable LISA sources. We present a
preliminary study of LISA's sensitivity to MBHB parameters using a
hybrid numerical/analytic waveform for equal-mass, non-spinning holes.
The \emph{Synthetic LISA} software package is used to simulate the
instrument response and the Fisher information matrix method is used
to estimate errors in the parameters. Initial results indicate that
inclusion of the merger signal can significantly improve the precision
of some parameter estimates. For example, the median parameter errors
for an ensemble of systems with total redshifted mass of $10^{6}\,\mbox{M}_{\odot}$
at a redshift of $z\sim1$ were found to decrease by a factor of slightly
more than two for signals with merger as compared to signals truncated
at the Schwarzchild ISCO.
\end{abstract}

\section{Introduction\label{sec:Introduction}}

The inspiral and merger of massive black hole binaries (MBHBs) are
expected to produce the strongest gravitational wave (GW) signals
in the LISA frequency band. The large signal-to-noise ratios (SNRs)
of these signals will allow LISA to not only detect, but measure physical
parameters of the source systems. This capability will allow LISA
to probe a wealth of astrophysical questions\cite{Science_Case} including
testing models of hierarchical structure formation through MBHB merger
trees\cite{Structure_Formation} and constraining cosmology through
direct measurements of the distance-redshift relationship at high
redshifts\cite{Standard_Sirens}.

Estimates of the precision with which LISA can extract waveform parameters
have been developing for the past several years, in parallel with
progress in describing MBHB waveforms. Initial studies focused on
equal-mass, non-spinning holes\cite{Cutler_Resolution}, while more
recent analyses have added the effects of spin\cite{Vecchio_Spin,LangHughes_Spin}
and higher harmonics\cite{Arun_Harmonics,PorterCornish_Harmonics}.
All of these works focus on the inspiral phase of the MBHB waveform,
truncating the signals at some time before the final merger and ringdown
occur. 

This truncation is justified given the mechanism by which the bulk
of LISA's parameter sensitivity is expected to arise, namely modulations
imparted on the signal from the orbital motion of the LISA constellation\cite{Cutler_Resolution}.
Although the merger signal contributes a significant fraction of SNR,
it does so in such a brief period of time that the orbital modulations
have little effect on the signal, reducing its usefulness for parameter
estimation. The other, more practical reason why the merger has been
excluded is that the post-Newtonian (PN) approximations used to compute
the waveforms break down at or near the merger and are not expected
to give accurate descriptions of the merger waveform. 

Numerical relativity (NR) has now provided a description of the complete
waveform from inspiral through merger and ringdown\cite{GSFC_Waveform,Texas_Waveform},
at least for a restricted set of parameters. Analyses using hybrid
PN/NR waveforms confirm that the merger portion of the signal often
dominates the total SNR of a LISA MBHB observation\cite{GSFC_GWobs}.
These waveforms make it possible to check the assumption that the
merger has a minimal effect on parameter estimation\cite{Thorpe_AAS08,Babak_PE}.
In this paper we present an initial study of LISA parameter estimation
using complete waveforms for equal-mass, non-spinning MBHBs.

\section{Methodology\label{sec:Methodology}}

\subsection{Waveform\label{sub:Waveform}}

The waveform used in these parameter estimation studies is a hybrid
of PN and NR waveforms as described in section IV of Baker, \etal.\cite{GSFC_GWobs}.
Briefly, the PN waveform (3.5PN in phase and 2.5PN in amplitude) spans
the time from $2\times10^{5}\, M_{rest}(G/c^{3})$ to $328M_{rest}(G/c^{3})$
before the merger, where $M_{rest}$ is the total mass of the MBHB.
At $328M_{rest}(G/c^{3})$, the waveform is stitched to a NR simulation
that extends through merger and ringdown. The matching time is chosen
at the point where the estimated accuracy of the NR simulations exceed
those of the PN expressions.

Since the hybrid waveform relies on NR simulations to generate the
merger segment, it is not practical to use this technique to generate
the large sets of waveforms with varying parameters that are needed
for many data analysis techniques. However, if we restrict ourselves to variations
in extrinsic parameters and keep intrinsic parameters such as mass ratio and
hole spins fixed, it is possible to use a single {}``master''
waveform and vary the extrinsic parameters using analytic formulae. 

For our purposes, the desired waveforms are dimensionless strain in
the Solar System Barycenter (SSB) frame, $h_{SSB}$. These depend on six extrinsic
parameters: the redshifted total system mass $M=M_{rest}(1+z)$, luminosity
distance $D_{L}$, coalescence time $t_{c}$, and three angles describing
the orientation of the binary, in this case an inclination $\iota$,
initial orbital phase $\phi_{o}$, and a polarization $\psi$.  It is convenient
to construct a rescaled complex strain in the source frame, 
$\hc\equiv R_{ext}\left(h_{+}+ih_{\times}\right)$,
which scales with $M_{rest}$, has a constant amplitude in the far-field limit,
and can accommodate the necessary angular dependencies through rotations in the
complex plane.  $R_{ext}$ is a conventional notation for the radius at which
the radiation has been extracted in the numerical simulation, although in our case, 
it can equally well be used as a point of reference for rescaling the PN amplitude. 
With these definitions, we
can use the following relationship between the rescaled strain in the source
frame and the dimensionless strain in the SSB, $h_{SSB}$:
\begin{equation}
h_{SSB}=\sqrt{\frac{5}{4\pi}}\frac{G}{c^{2}}\frac{M}{D_{L}}e^{2i\psi}\left[\hc\cos^{4}\left(\frac{\iota}{2}\right)e^{2i\phi_{o}}+\hc^{*}\sin^{4}\left(\frac{\iota}{2}\right)e^{-2i\phi_{o}}\right]\label{eq:source2SSB}
\end{equation}

Two additional extrinsic parameters describe the sky location of the
binary in the SSB frame. In this case we use galactic longitude $\Phi$
and latitude $\Theta$. The effect of these parameters of the waveform
is handled by the instrument model, described in the next section.
In total, our model uses eight extrinsic parameters to determine $h_{SSB}$
from $\hc$. In the remainder of this work, the vector
$\vec{\lambda}\equiv(M,D_{L},\Theta,\Phi,\iota,\phi_{o},\psi,t_{c})$
denotes these parameters and the shorthand notation $h = h(\vec{\lambda})$
represents the transformation in (\ref{eq:source2SSB}) .

\subsection{Instrument Model\label{sub:InstrumentModel}}

There are two necessary components for modeling the sensitivity of
a gravitational wave detector\cite{Larson_Sensitivity}: a response
function relating the gravitational wave signal to the instrument
output and a noise model describing all other components of the instrument
output. 

In the case of LISA, the instrument outputs are the Time Delay Interferometry\cite{TDIOriginal,Krolak_TDI}
(TDI) streams. TDI is a technique by which the differential phase
measurements made at each spacecraft are combined with appropriate
time delays in order to form outputs for which laser phase noise is
canceled but GW signals remain. For data analysis purposes, it is
convenient to use a set of {}``optimal'' TDI channels\cite{JPL_Optimal_TDI}
that form an orthogonal basis in the TDI signal space. For an optimal
set, each channel contains independent noise and can therefore be
treated as a separate detector. We have defined a set of optimal channels
$\mathbf{A}\equiv(A^{\prime},E^{\prime},T^{\prime})$ where $A^{\prime}=\frac{1}{2}\frac{1}{\sqrt{2}}(Z-X)$,
$E^{\prime}=\frac{1}{2}\frac{1}{\sqrt{6}}(Z+X-2\cdot Y)$, and $T^{\prime}=\frac{1}{2}\frac{1}{\sqrt{3}}(Z+X+Y)$
where $(X,Y,Z)$ are the first-generation Michelson TDI variables.
The primed notation is used to distinguish $\mathbf{A}$ from the
$(A,E,T)$ used by Prince, \etal.\cite{JPL_Optimal_TDI} .

The \textit{Synthetic LISA} software package\cite{SynthLISA} is used
to model the instrument response to the GW signal. \textit{Synthetic
LISA} models the orbits of the individual spacecraft and computes
the constellation geometry. The GW response in each arm is then computed
and assembled to form the appropriate TDI combinations. Although the
computational cost of a numerical simulation exceeds that of an analytic
model, the \textit{Synthetic LISA} \textit{\emph{model provides higher
fidelity, particularly at high Fourier frequencies where the instrument
response becomes complex and time-dependent.}}

It would also be possible to use \textit{Synthetic LISA} to compute
the noise output in the $\mathbf{A}$ channels as well as the signal
response. The difficulty with this approach is that statistical variations
cause the noise curve to change from one run to the next, washing
out the subtle changes corresponding to variations in waveform parameters.
This effect could be mitigated by averaging repeated simulations with
varying seeds, however this would add to the computational cost.

An alternative approach, the one taken by most estimates of LISA sensitivity
in the literature, is to derive an analytic expression for the average
noise level in a TDI channel given the noise spectra of the individual
components. For $\mathbf{A}$, the one-sided power-spectral densities
(PSDs) of the noise are given by
\begin{eqnarray}
S_{A^{\prime},E^{\prime}} &=& 2\sin^{2}\left(\zeta\right)\cdot\left\{ 2\cdot\left[3+2\cos\left(\zeta\right)+\cos\left(2\zeta\right)\right]S_{pm}(f) \right. \nonumber \\
   && \left. +\left[2+\cos\left(\zeta\right)\right]S_{op}(f)\right\} ,\label{eq:AE_Noise}\\
S_{T^{\prime}}&=&8\sin^{2}\left(\zeta\right)\cdot\sin^{2}\left(\zeta/2\right)\cdot\left[4\sin^{2}\left(\zeta/2\right)S_{pm}(f)+S_{op}(f)\right],\label{eq:T_Noise}\end{eqnarray}
where $\zeta\equiv2\pi fL/c$, $f$ is the Fourier frequency, $L$
is the average detector arm-length, and $S_{pm}(f)$ and $S_{op}(f)$
are the strain-equivalent proof-mass acceleration and optical-path
noises of the LISA links respectively. For this work, $S_{pm}(f)=\left(2.5\times10^{-48}\,\mbox{Hz}^{-1}\right)\cdot\left(f/1\,\mbox{Hz}\right)^{-2}\cdot\sqrt{1+\left(f/0.1\,\mbox{mHz}\right)^{-4}}$
and $S_{op}(f)=\left(1.8\times10^{-37}\,\mbox{Hz}^{-1}\right)\cdot\left(f/1\,\mbox{Hz}\right)^{2}$.

In addition to the noise introduced by the instrument, MBHB signals
are also partially obscured by a {}``foreground'' of other GW signals.
This foreground is composed of a superposition of tens of thousands
of compact binaries in our galaxy and its neighborhood. We estimate
the noise contribution in $A^{\prime}$ and $E^{\prime}$ of this
stochastic foreground as $S_{gal}(f)=\left[4\zeta\sin\left(\zeta\right)\right]^{2}\cdot S_{conf}(f)$,
where $S_{conf}(f)$ is taken from equation (14) of Timpano, Rubbo,
and Cornish\cite{Galactic_Background}. The pre-factor in $S_{gal}(f)$
is an estimate of the signal response functions of $A^{\prime}$ and
$E^{\prime}$ for the frequencies where $S_{conf}(f)$ is defined,
$40\,\mu\mbox{Hz}\leq f\leq8\,\mbox{mHz}$. At these low frequencies,
the response in the $T^{'}$ channel is near zero. Consequently no
foreground is added to the $T^{\prime}$ channel. 

One downside to using different instrument models for the signal and
the noise is that artifacts can arise when these models are inconsistent.
For example, the analytic formulas predict zero noise at $f_{n}=nc/L,\: n=1,2,3...$.
Were analogous formulae used to predict the signal response, the result
would be zero signal at $f=f_{n}$. In practice, the \textit{Synthetic
LISA} model predicts some finite signal at $f=f_{n}$, partly due
to errors in estimating PSDs from discrete time series. The result
is an infinite contribution to the SNR at $f=f_{n}$. To alleviate
this problem, we added an artificial {}``floor'' of $\left(10^{-40}\,\mbox{Hz}^{-1}\right)\cdot(f/1\,\mbox{Hz})^{2}$
to the noise models for $\mathbf{A}$. 

The $T^{\prime}$ channel has a related problem at low frequencies,
where $S_{T^{\prime}}\propto f^{2}$. The corresponding signal response
function should also fall off with $f^{2}$ at low frequencies in
order for SNR to converge. In practice, the $T^{\prime}$ channel
response for our signals leveled off at a finite value, leading to
an erroneous increase in SNR (and parameter sensitivity). It may be
possible to address the problem with the $T^{\prime}$ channel by
adding an artificial noise floor to $S_{T^{\prime}}(f)$ or by only
using it in the high-frequency limit. For this work, however, we have
chosen to neglect the contributions of the $T^{\prime}$ channel until
the issue can be further investigated.

Finally, we note that these types of artifacts may arise when actual
instrumental data is analyzed. Most analysis methods require some
model of the instrument response and noise, each of which will have
some discrepancies with the actual instrument behavior.

\subsection{Parameter Sensitivity Calculation\label{sub:ParameterCalc}}

With the waveform described in Sec. \ref{sub:Waveform} and the instrument
model described in Sec. \ref{sub:InstrumentModel} in hand, the next
step is to develop a procedure for evaluating parameter sensitivity.
Like many others in the community, we have selected the Fisher information
matrix approach\cite{Vallisneri_Fisher}. 

The steps for the approach are as follows. A SSB waveform with a particular
set of extrinsic parameters $\vec{\lambda}_{o}$ is generated from
the master waveform, $h_{o}=h_{SSB}\left(\vec{\lambda}_{o}\right)$.
A set of perturbed waveforms is then generated by perturbing each
of the parameters individually, $h_{i}\equiv h_{SSB}(\vec{\lambda}_{o}+\delta\lambda_{i}\,\hat{\imath})$,
where $\delta\lambda_{i}$ is the size of the perturbation in the
$\hat{\imath}$ parameter direction. The instrument response to $h_{o}$
and $h_{i}$ is computed using \emph{Synthetic LISA}. The
partial derivative of the instrument response with respect to each
parameter is then estimated using a one-sided difference,
\begin{equation}
\left.\frac{\partial\mathbf{A}}{\partial\lambda^{i}}\right|_{\vec{\lambda}=\vec{\lambda}_{o}}\approx\frac{\mathbf{A}_{i}-\mathbf{A}_{o}}{\delta\lambda_{i}}\label{eq:paramDer}
\end{equation}
where $\mathbf{A}_{o}$ is the response to $h_{o}$ and $\mathbf{A}_{i}$
is the response to $h_{i}$. These parameter derivatives are
used to estimate the Fisher information matrix
\begin{equation}
\mathbf{\Gamma}_{ij}=\left(\frac{\partial\mathbf{A}}{\partial\lambda^{i}}\left|\frac{\partial\mathbf{A}}{\partial\lambda^{j}}\right.\right)\label{eq:fishDef}
\end{equation}
where $(a|b)$ denotes the noise-weighted inner product,
\begin{equation}
(a|b)\equiv2\int_{0}^{\infty}\frac{a^{\star}(f)b(f)+a(f)b^{\star}(f)}{S_{n}(f)}df.\label{eq:InnerPdef}
\end{equation}
The quantity $S_{n}(f)$ is the PSD of the appropriate TDI channel.
In our case we estimate the Fourier transform of $\partial\mathbf{A}/\partial\lambda^{i}$
from the time series using a discrete Fourier transform with tapering
applied to the early and late portion of the signal to reduce spectral
artifacts. The inner product in (\ref{eq:InnerPdef}) is evaluated
as a discrete sum with finite upper and lower integration limits.

Since the $\mathbf{A}$ channels are constructed to be linearly independent,
it is appropriate to sum their individual Fisher matrices to yield
the effective Fisher matrix for the instrument as a whole. However,
as discussed in section \ref{sub:InstrumentModel}, the $T^{\prime}$
channel was not well-behaved in our instrument model. For that reason,
we have omitted the $T^{\prime}$ channel and defined the composite
Fisher matrix as $\Gamma_{ij}=\Gamma_{ij}^{(A^{\prime})}+\Gamma_{ij}^{(E^{\prime})}$.
Taking the usual limit of large SNR, we compute the covariance matrix
$\Sigma^{ij}\approx\left(\Gamma^{-1}\right)^{ij}$ and parameter errors
$\Delta\lambda^{i}=\sqrt{\Sigma^{ij}}$. In order to explore the effect
of the choice of $\vec{\lambda}_{o}$, we perform multiple runs with
varying $\vec{\lambda}_{o}$ and estimate the probability density
functions of $\Sigma^{ij}$ and $\Delta\lambda^{i}$.

\section{Results\label{sec:Results}}

\begin{figure}
\begin{centering}\subfigure[$SNR_{AE}$]{\includegraphics[width=4cm,keepaspectratio]{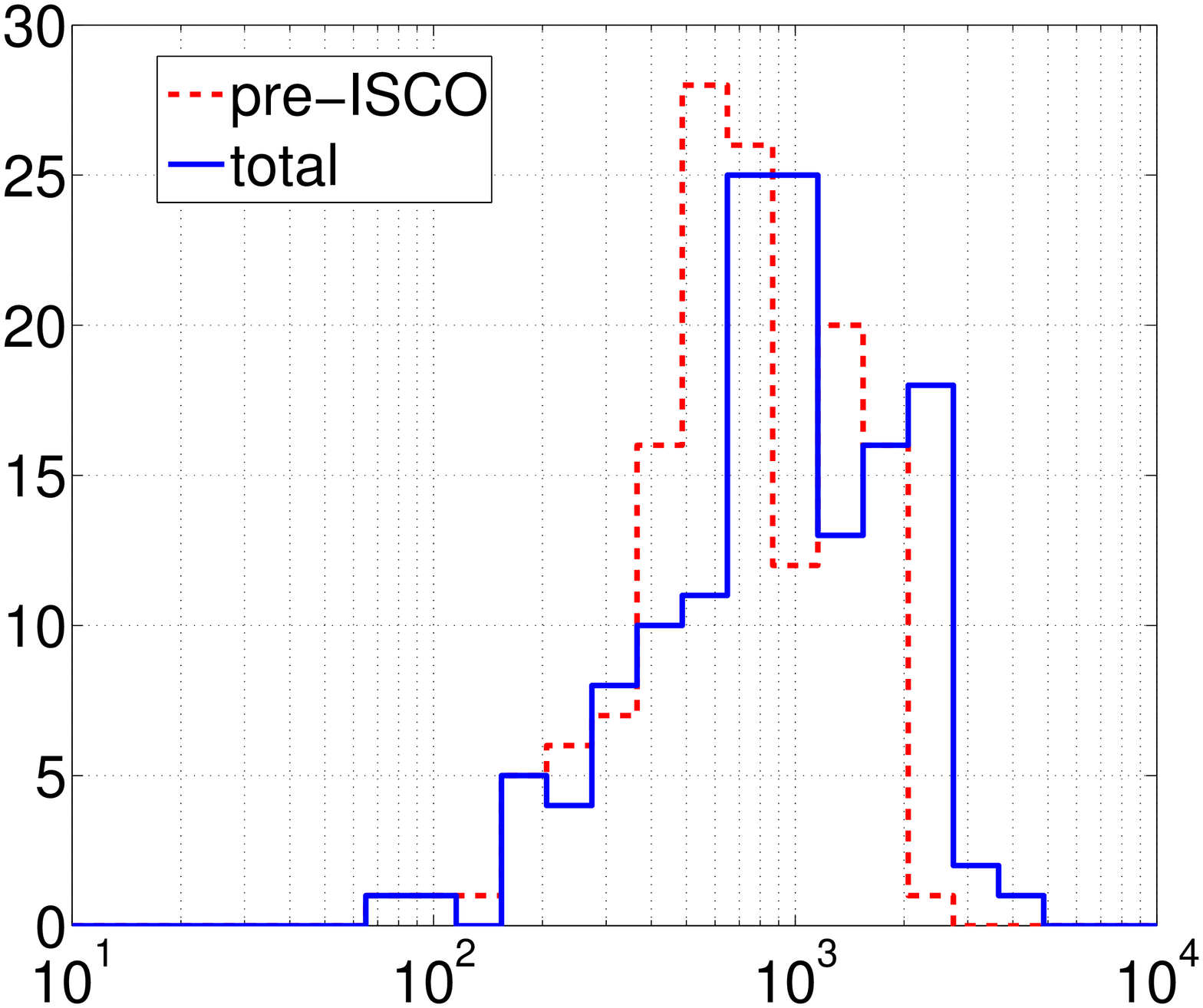}}\quad\subfigure[$\Delta\ln\left(M\right)$]{\includegraphics[width=4cm,keepaspectratio]{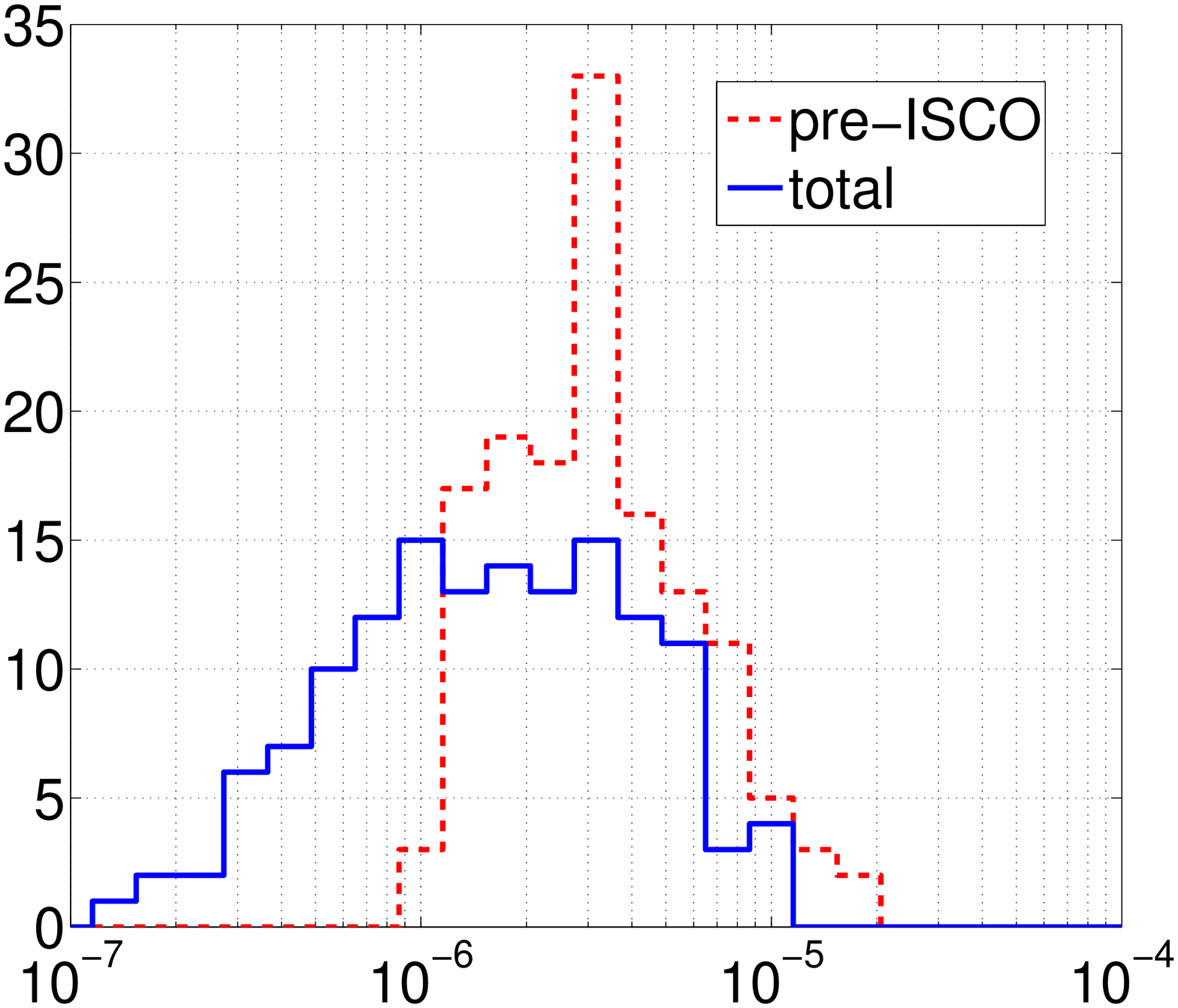}}\quad\subfigure[$\ln\left(D_{L}\right)$]{\includegraphics[width=4cm,keepaspectratio]{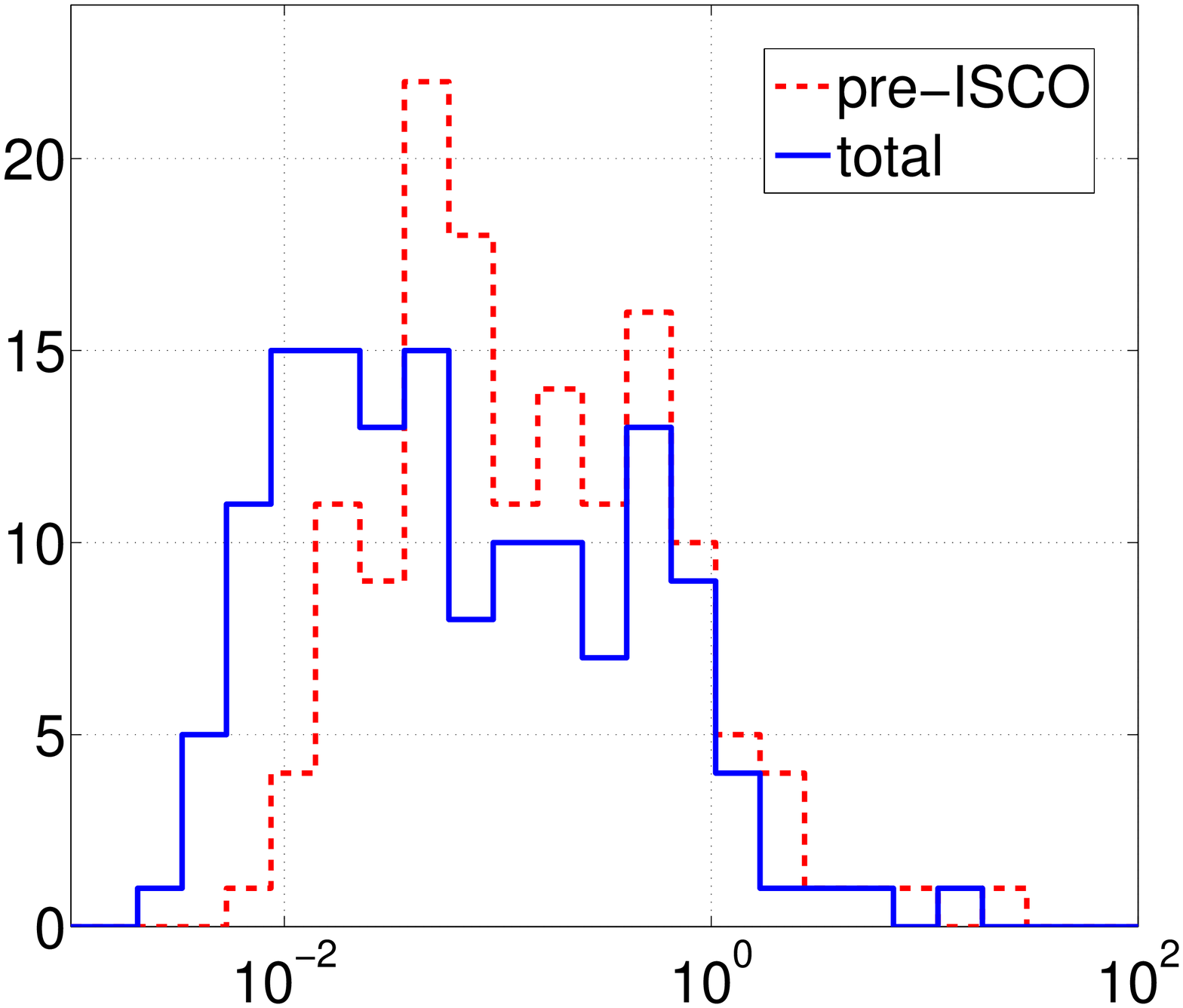}}\\
\subfigure[$\Delta\Theta$ (deg)]{\includegraphics[width=4cm,keepaspectratio]{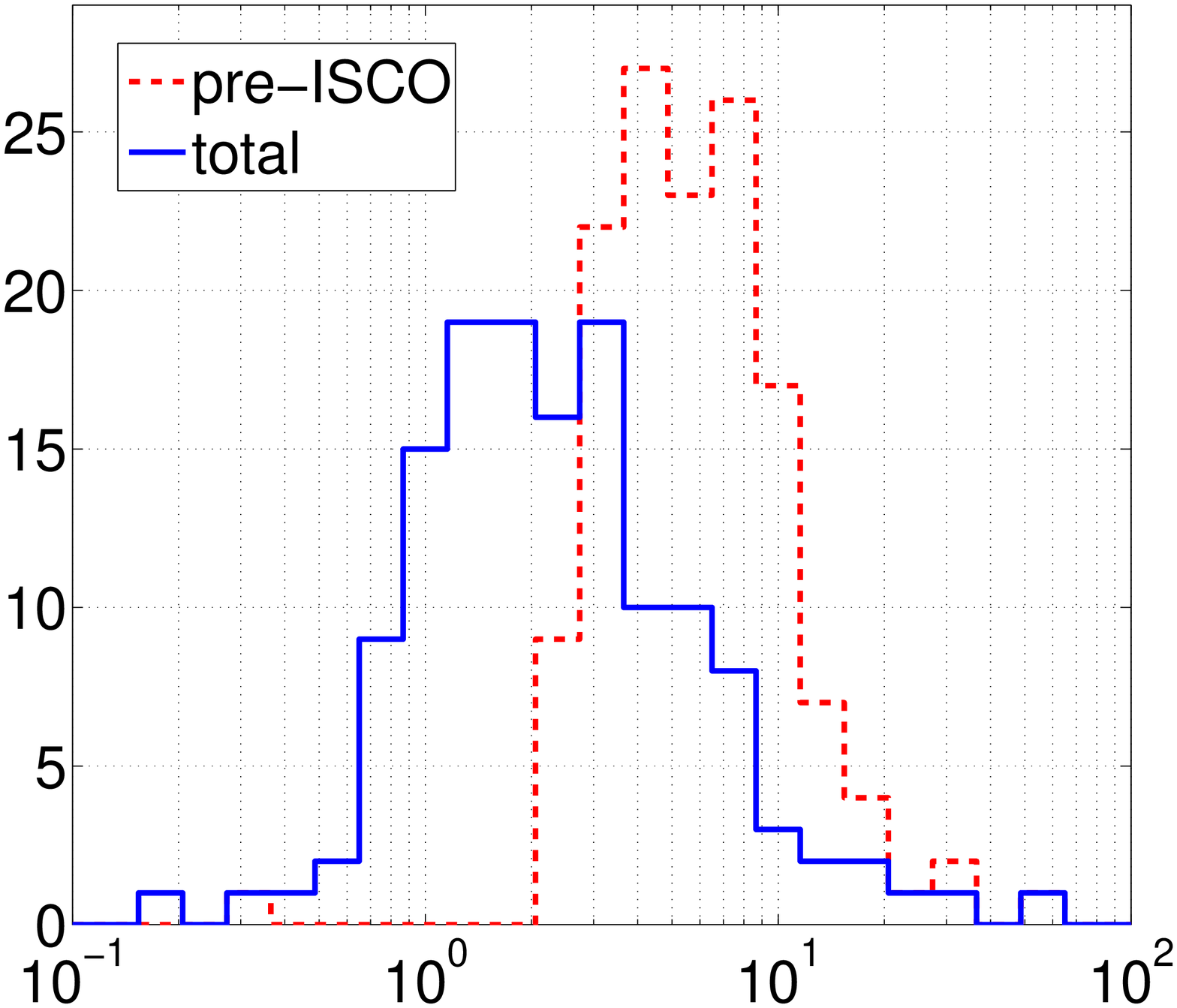}}\quad\subfigure[$\Delta\Phi$ (deg)]{\includegraphics[width=4cm,keepaspectratio]{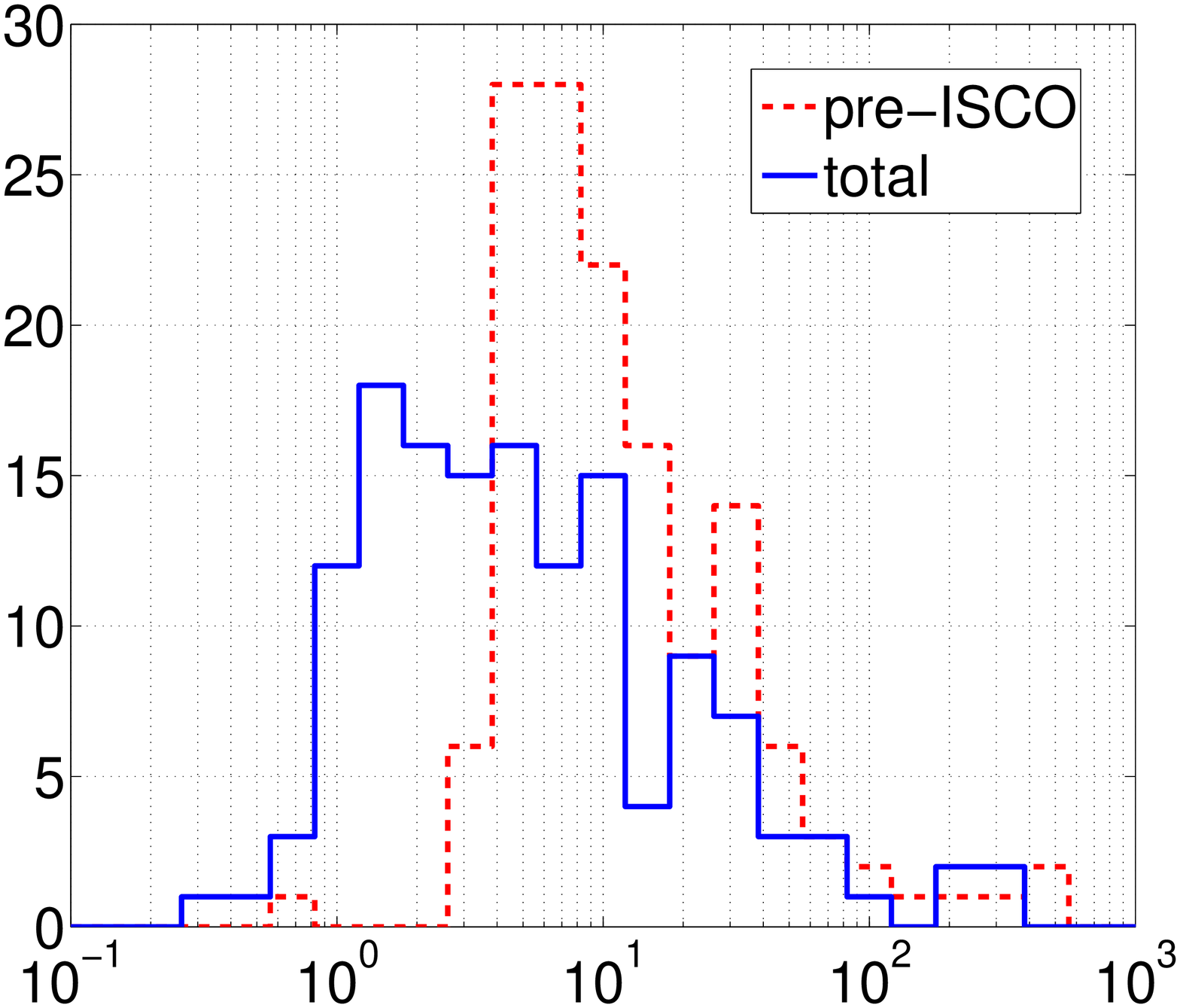}}\quad\subfigure[$\Delta\iota$ (deg)]{\includegraphics[width=4cm,keepaspectratio]{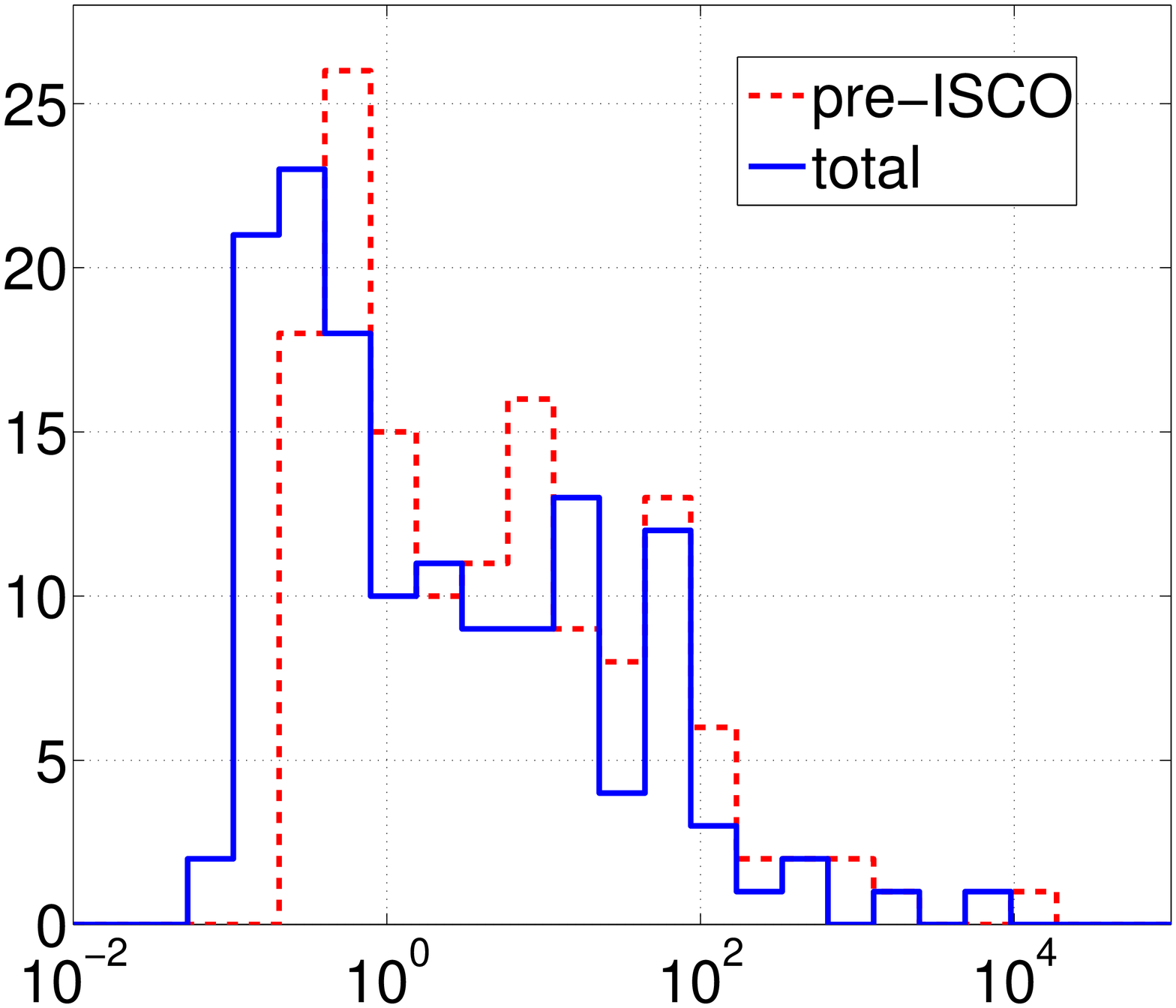}}\\
\subfigure[$\Delta\phi_{o}$ (deg)]{\includegraphics[width=4cm,keepaspectratio]{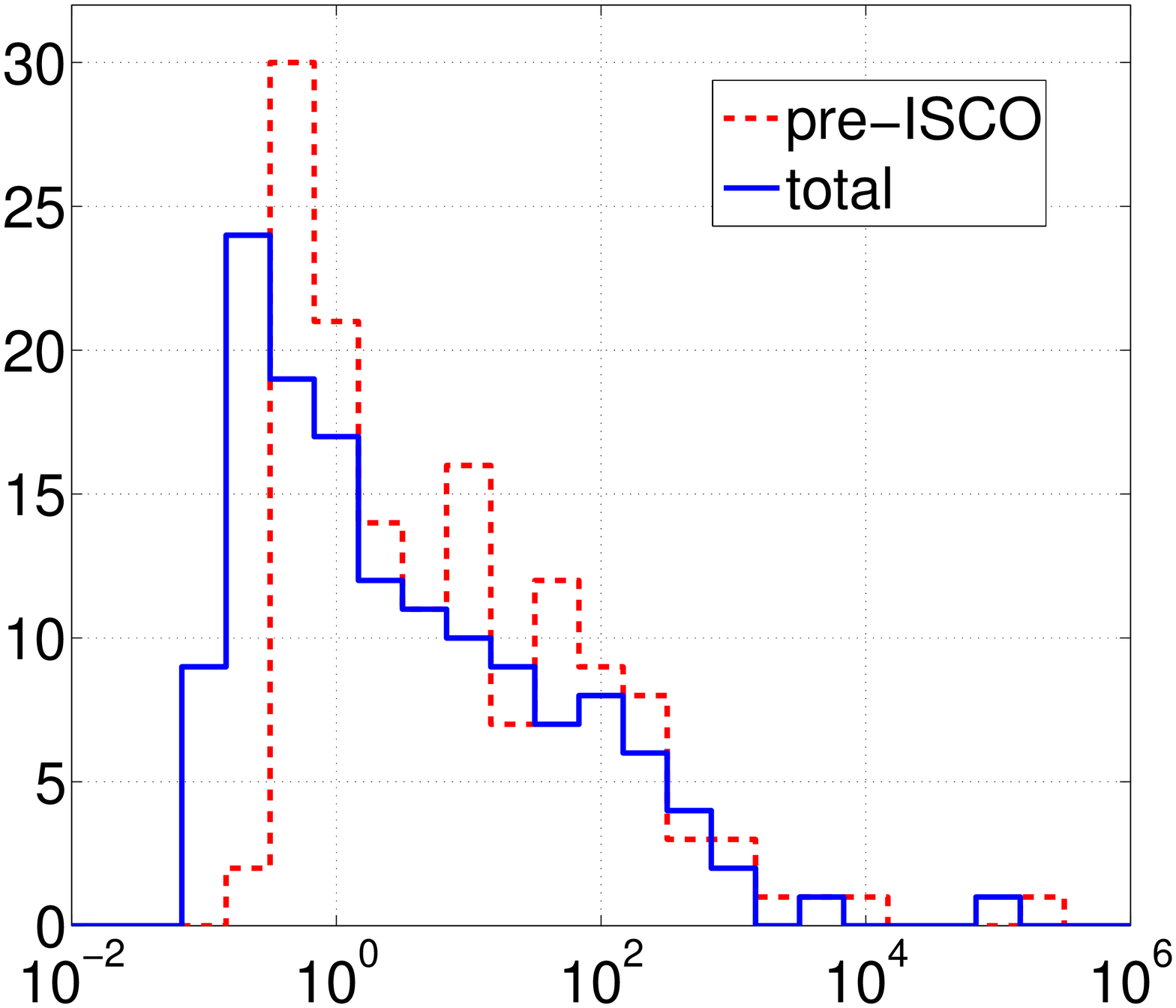}}\quad\subfigure[$\Delta\psi$ (deg)]{\includegraphics[width=4cm,keepaspectratio]{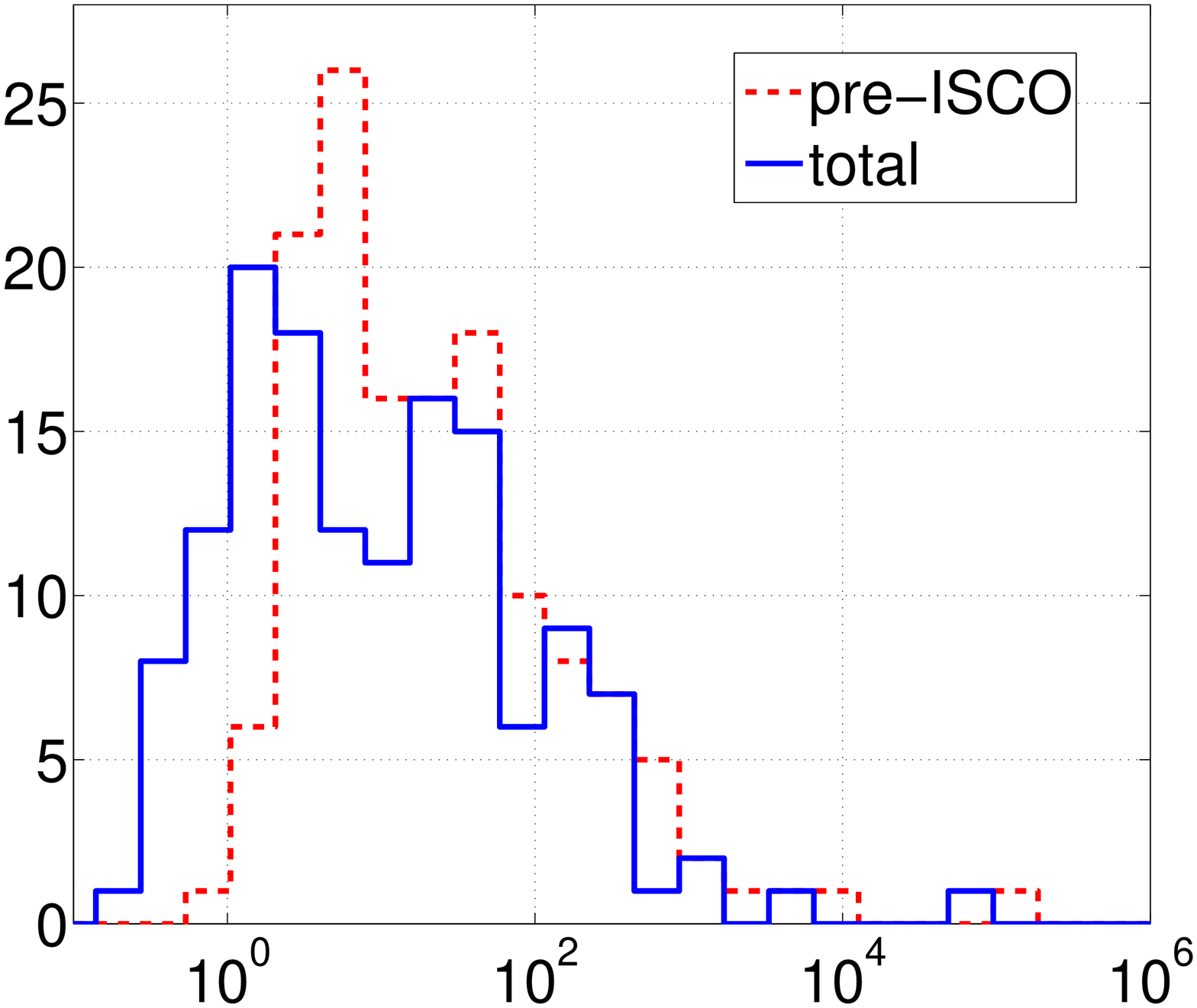}}\quad\subfigure[$\Delta t_{c}$ (sec)]{\includegraphics[width=4cm,keepaspectratio]{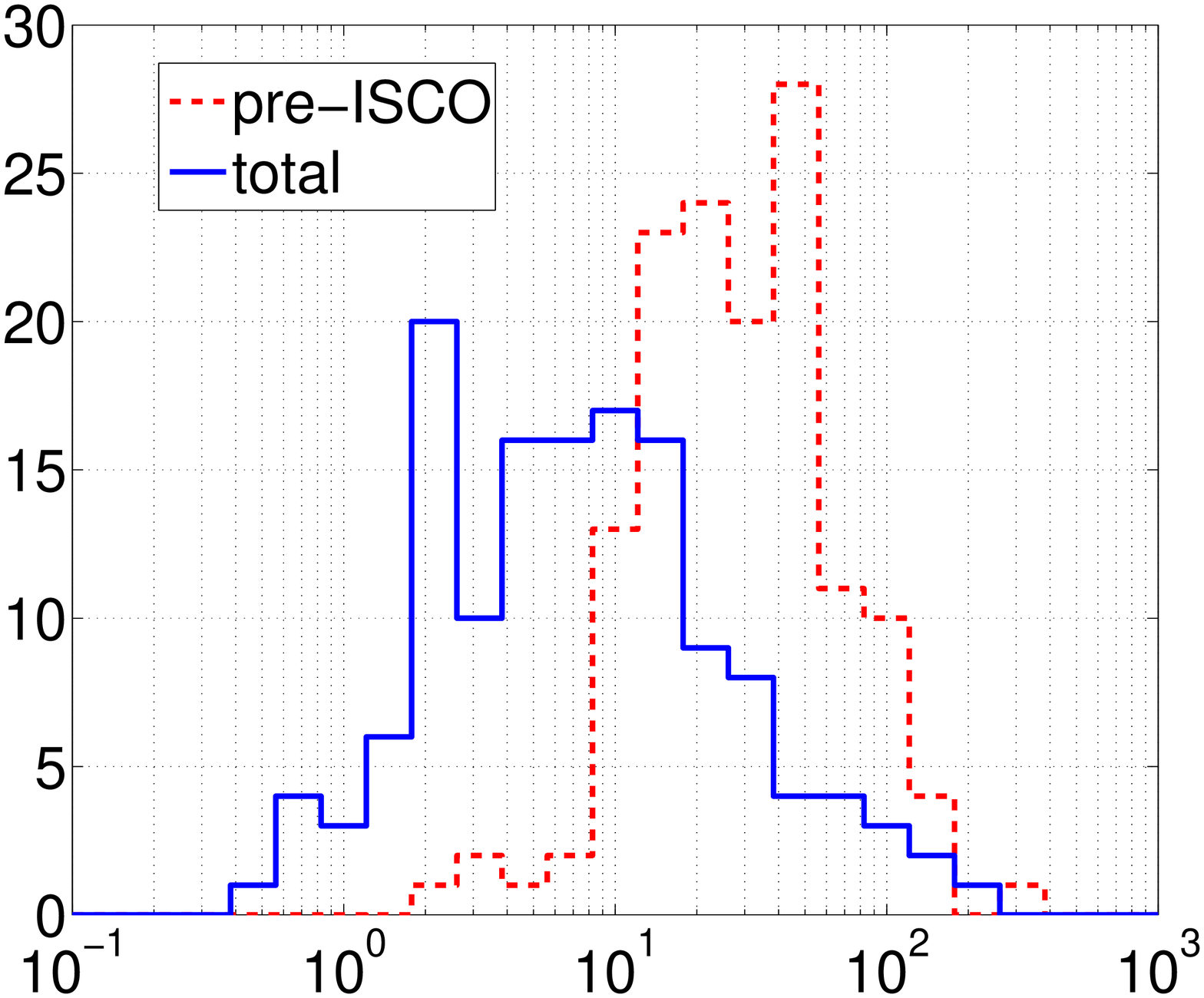}}\par\end{centering}

\caption{\label{fig:M6hists}Histograms of SNR and $\Delta\lambda^{i}$ for
ensemble summarized in Table \ref{tab:M6summary}. The dashed lines
are the results with a cutoff at ISCO, the solid lines include the
entire waveform.}
\end{figure}

For our initial trials, we choose an ensemble of 140 systems, each
with $M=10^{6}\,\mbox{M}_{\odot}$ and $D_{L}=6\,\mbox{Gpc}$, corresponding
to $z\sim1$ in standard $\Lambda$-CDM cosmology. The remaining parameters
($\Theta$, $\Phi$, $\iota$, $\phi_{o}$, $\psi$, and $t_{c}$)
were randomly varied over the ensemble. For each member of the ensemble,
$\Gamma_{ij}$ was computed using two different upper frequency limits
for the inner product in (\ref{eq:InnerPdef}). In one case, the integral
was truncated at the Schwarzchild inner-most stable circular orbit
(ISCO) frequency whereas in the second case it was continued for an
order of magnitude beyond the peak (merger) frequency. Table \ref{tab:M6summary}
lists the mean and median values of the SNR and parameter errors for
the ensemble. Histograms of each parameter error for both cases are
shown in Figure \ref{fig:M6hists}. Note that the plotted quantities are 
the direct results of the technique outlined in \ref{sec:Methodology}, without 
imposing any physical limits on the sizes of the errors. For example, an error
in galactic latitude of $\sim1000\,\mbox{deg}$ should be interpreted as the 
measurement providing no useful $\Theta$ information for that system.

\begin{table}

\caption{\label{tab:M6summary}Mean and median values for parameter errors
and SNR for an ensemble of 140 equal-mass, non-spinning binaries with
$M=10^{6}\,\mbox{M}_{\odot}$ at $D_{L}=6\,\mbox{Gpc}$ ($z\sim1)$.
For 'pre-ISCO' , integrations of Fisher matrix elements were cut off
at a frequency corresponding to Schwarzchild ISCO. }

\begin{centering}\begin{tabular}{|c|c|c|c|c|}
\hline 
&
total mean&
pre-ISCO mean&
total median&
pre-ISCO median\tabularnewline
\hline
\hline 
$SNR_{AE}$&
$1.12\times10^{3}$&
$8.35\times10^{2}$&
$9.28\times10^{2}$&
$6.98\times10^{2}$\tabularnewline
\hline 
$\Delta\ln(M/M_{\odot})$&
$2.35\times10^{-6}$&
$4.00\times10^{-6}$&
$1.66\times10^{-6}$&
$3.26\times10^{-6}$\tabularnewline
\hline 
$\Delta\ln(D_{L}/1\,\mbox{pc})$&
$3.78\times10^{-1}$&
$6.07\times10^{-1}$&
$4.48\times10^{-2}$&
$1.18\times10^{-1}$\tabularnewline
\hline 
$\Delta\Theta$ (deg)&
$3.84$&
7.19&
2.15&
5.28\tabularnewline
\hline 
$\Delta\Phi$ (deg)&
17.7&
26.9&
4.64&
9.19\tabularnewline
\hline 
$\Delta\iota$ (deg)&
81.9&
137&
1.08&
3.12\tabularnewline
\hline 
$\Delta\phi_{o}$ (deg)&
607&
1270&
1.59&
4.26\tabularnewline
\hline 
$\Delta\psi$ (deg)&
620&
1290&
7.46&
15.9\tabularnewline
\hline 
$\Delta t_{c}$(sec)&
17.1&
39.5&
6.87&
27.3\tabularnewline
\hline
\end{tabular}\par\end{centering}
\end{table}

The median improvement in $\Delta\lambda^{i}$ upon inclusion of the
merger is a factor of $\sim2$, with the exception of $\Delta t_{c}$,
which improves by a factor of $\sim4$. In all cases, the improvement
factor is greater than that for the median SNR of $\sim$25\%, suggesting
that most of the improvement is via a mechanism other than direct
increase in SNR. 

One possible mechanism is that a pair of parameters that is nearly
degenerate in the inspiral phase becomes decoupled in the more complex
merger phase of the waveform. Table \ref{tab:M6_Cov_Decrease} lists
the percentage decrease in covariance for each parameter pair from
pre-ISCO to total waveform. Note that all of the covariances decrease
when the post-ISCO waveform is added. Covariances with the coalescence
time exhibit the most dramatic decrease. This can be qualitatively
explained by arguing that the merger provides a sharp feature that
constrains the coalescence time more tightly and breaks degeneracies
with other parameters.

\begin{table}

\caption{\label{tab:M6_Cov_Decrease}Median percentage reduction in covariance
from pre-ISCO to total signal, $1-\Sigma_{ij}^{(total)}/\Sigma_{ij}^{(pre-ISCO)}$.}

\begin{centering}\begin{tabular}{|c|c|c|c|c|c|c|c|c|}
\hline 
&
$\ln M$&
$\ln D_{L}$&
$\Theta$&
$\Phi$&
$\iota$&
$\phi_{o}$&
$\psi$&
$t_{c}$\tabularnewline
\hline
\hline 
$\ln M$&
72.3&
70.8&
63.5&
90.7&
71.4&
79.2&
82.4&
91.1\tabularnewline
\hline 
$\ln D_{L}$&
-&
75.7&
79.6&
82.3&
76.4&
83.5&
83.4&
95.4\tabularnewline
\hline 
$\Theta$&
-&
-&
80.9&
89.6&
80.8&
77.2&
81.9&
97.6\tabularnewline
\hline 
$\Phi$&
-&
-&
-&
82.9&
83.1&
80.8&
81.8&
87.2\tabularnewline
\hline 
$\iota$&
-&
-&
-&
-&
76.8&
87.0&
86.6&
94.8\tabularnewline
\hline 
$\phi_{o}$&
-&
-&
-&
-&
-&
78.7&
78.5&
91.8\tabularnewline
\hline 
$\psi$&
-&
-&
-&
-&
-&
-&
78.3&
93.0\tabularnewline
\hline 
$t_{c}$&
-&
-&
-&
-&
-&
-&
-&
94.2\tabularnewline
\hline
\end{tabular}\par\end{centering}
\end{table}

\subsection{Evolution of Parameter Errors\label{sub:ErrorEvolution}}

The results in Table \ref{tab:M6summary}, Figure \ref{fig:M6hists},
and Table \ref{tab:M6_Cov_Decrease} demonstrate the improvement in
parameter estimation when the merger and ringdown portions of the
waveform are included. A related question is to ask how the parameter
information accumulates with time. This is of great interest for the
problem of searching for electromagnetic (EM) counterparts for MBHB
mergers. If the final portion of the signal provides significant improvement
in parameter estimates, there will be pressure to reduce the latency
between LISA and EM observatories. This could affect decisions on
data downlink operations, inter-spacecraft communications, and other
aspects of the mission.

\begin{figure}
\noindent \begin{centering}\subfigure[$SNR_{AE}$]{\includegraphics[width=5cm,keepaspectratio]{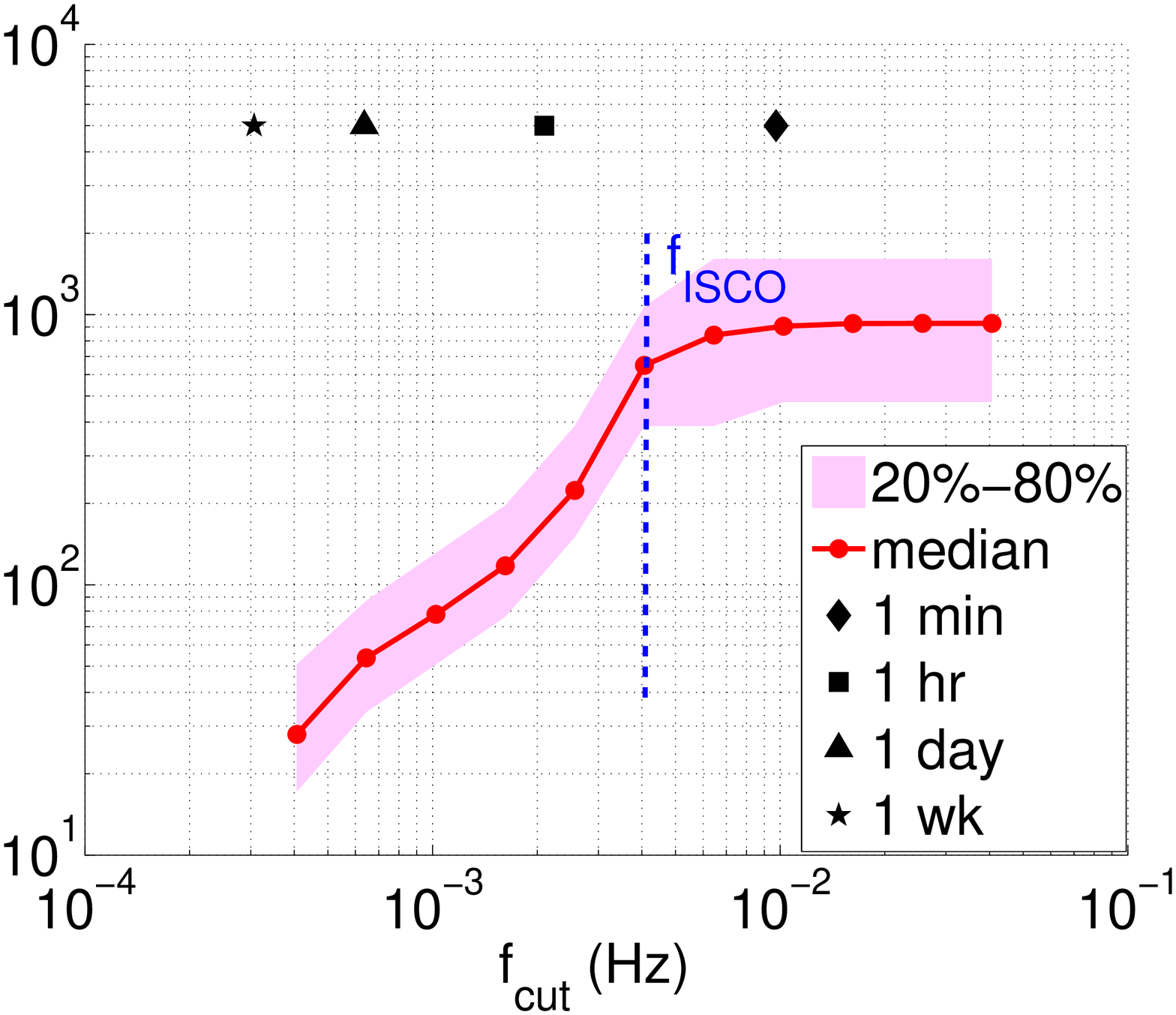}}\quad\subfigure[$\Delta \ln\left(D_{L}\right)$]{\includegraphics[width=5cm,keepaspectratio]{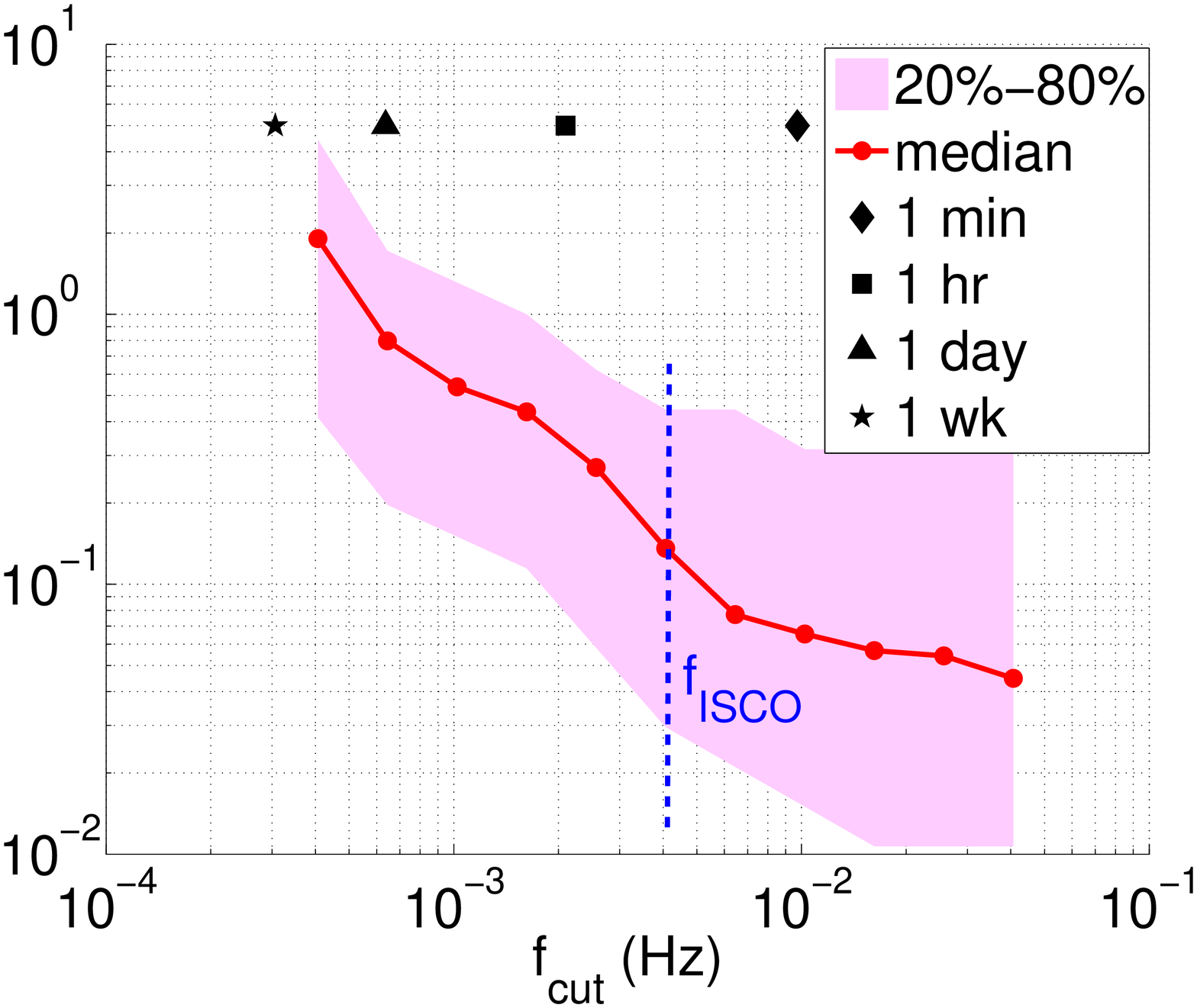}}\\
\subfigure[$\Delta\Theta$ ]{\includegraphics[width=5cm,keepaspectratio]{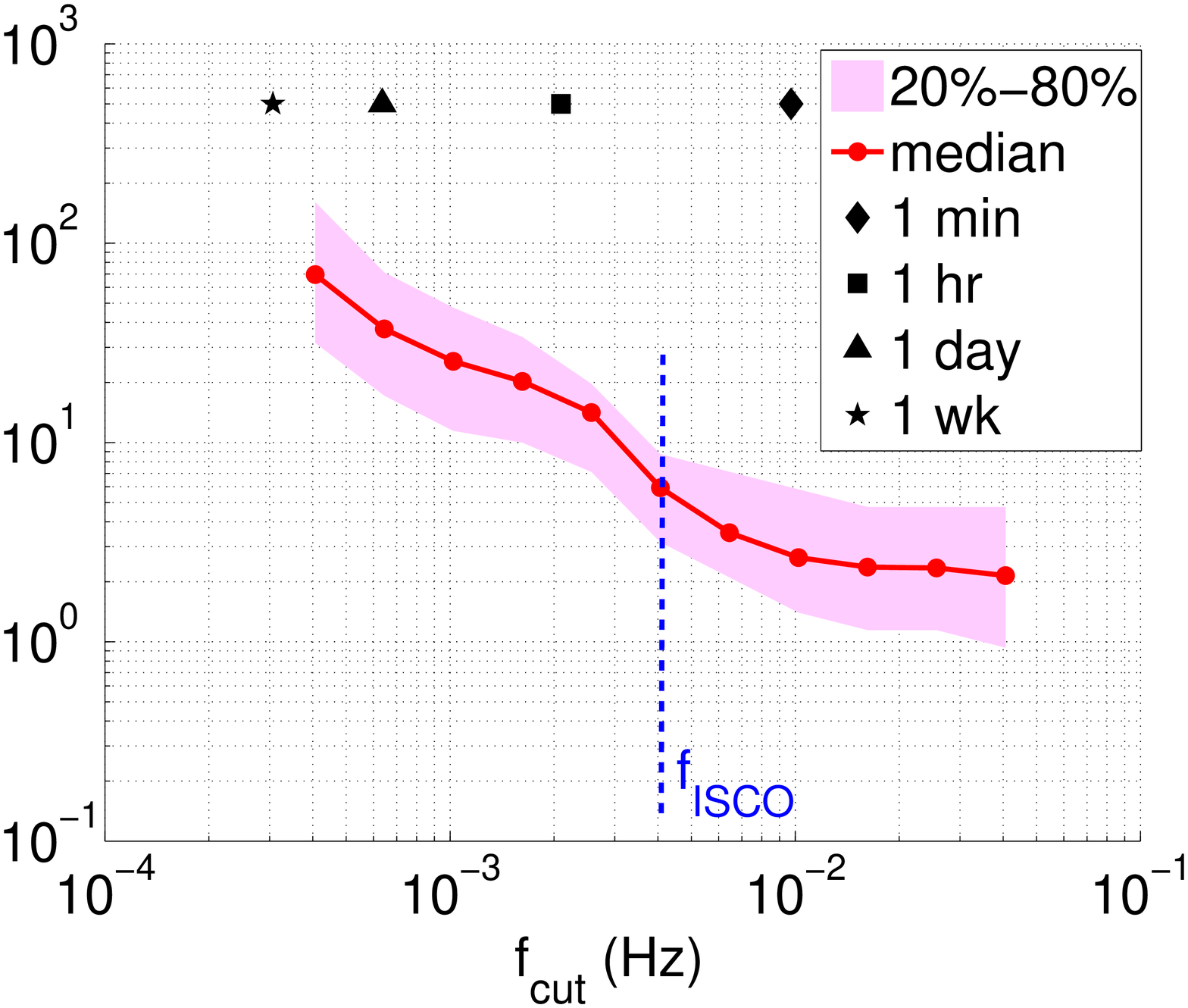}}\quad\subfigure[$\Delta\Phi$ ]{\includegraphics[width=5cm,keepaspectratio]{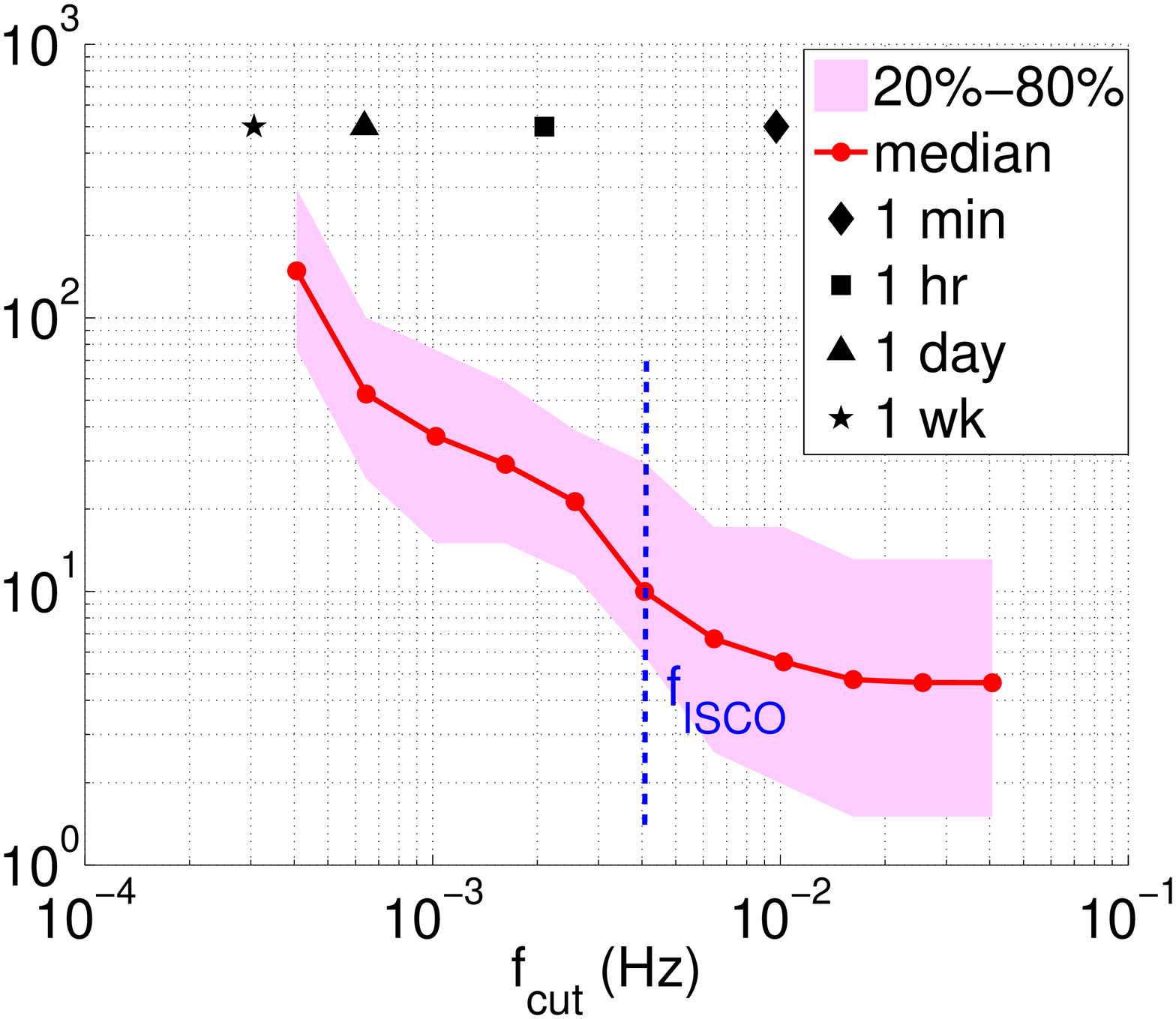}}\par\end{centering}

\caption{\label{fig:M6_upcut}Effect of upper frequency cut on SNR, $\Delta D_{L}$,
$\Delta\Theta$, and $\Delta\Phi$ for the ensemble summarized in
Table \ref{tab:M6summary}. Solid line denotes median values while
shaded spans $20^{th}$ to $80^{th}$ percentile range. The vertical
dashed line denotes the Schwarzchild ISCO frequency. Individual solid
markers denote time before merger.}
\end{figure}

Figure \ref{fig:M6_upcut} shows the effect of varying the upper cutoff
frequency of the inner product in (\ref{eq:InnerPdef}) on the SNR,
$\Delta D_{L}$, $\Delta\Theta$, and $\Delta\Phi$. From the EM counterpart
search perspective, these (along with perhaps $t_{c}$) are the most
critical parameters. Since our waveforms are composed of a single
mode, the GW frequency and ${(t-t}_{c})$ can be related analytically\cite{Blanchet_LivingReview}.
Included on the plots in Figure \ref{fig:M6_upcut} are points corresponding
to $(t-t_{c})=$ 1 min., 1 hr., 1 day, and 1 wk. Figure \ref{fig:M6hists}
indicates that $\Delta\Theta$ and $\Delta\Phi$ improve by more than
an order of magnitude in the final day before merger. Lang \& Hughes\cite{LangHughes_Localizing}
looked at spinning, precessing, unequal mass binaries using PN waveforms
and found that sky position error decreased by similar amount during
the final day before merger. 

\subsection{Effect of Mass}

As a test of the effect of total redshifted system mass on our parameter
estimation results, we simulated an ensemble of 140 equal-mass, non-spinning
MBHBs with $M=10^{7}\,\mbox{M}_{\odot}$ and $D_{L}=6\,\mbox{Gpc}$.
Table \ref{tab:M7summary} gives a summary of the mean and median
parameter errors for this run. Comparing with the analogous ensemble
of $M=10^{6}\,\mbox{M}_{\odot}$ systems summarized in Table \ref{tab:M6summary},
the heavier systems generally exhibit a reduced SNR and increased
parameter errors. This can be explained by the shift of the GW signal
to lower frequencies and its obscuration by the steeply-rising noise
of the LISA detector. As with the $M=10^{6}\,\mbox{M}_{\odot}$ systems,
the reduction in parameter errors with the inclusion of the merger
signal is still present in the $M=10^{7}\,\mbox{M}_{\odot}$ case.

\begin{table}

\caption{\label{tab:M7summary}Same as Table \ref{tab:M6summary} for an ensemble
of 140 equal-mass, non-spinning binaries with $M=10^{7}\,\mbox{M}_{\odot}$
at $D_{L}=6\,\mbox{Gpc}$ ($z\sim1)$. }

\begin{centering}\begin{tabular}{|c|c|c|c|c|}
\hline 
&
total mean&
pre-ISCO mean &
total median&
pre-ISCO median\tabularnewline
\hline
\hline 
$SNR_{AE}$&
98.5&
26.8&
85.2&
22.9\tabularnewline
\hline 
$\Delta\ln(M/M_{\odot})$&
$1.55\times10^{-6}$&
$2.66\times10^{-6}$&
$1.01\times10^{-6}$&
$1.89\times10^{-6}$\tabularnewline
\hline 
$\Delta\ln(D_{L}/1\,\mbox{pc})$&
2.17&
2.83&
$8.04\times10^{-1}$&
1.07\tabularnewline
\hline 
$\Delta\Theta$ (deg)&
22.4&
33.5&
19.8&
27.6\tabularnewline
\hline 
$\Delta\Phi$ (deg)&
58.4&
103&
22.0&
35.1\tabularnewline
\hline 
$\Delta\iota$ (deg)&
9.65&
25.2&
$5.88\times10^{-1}$&
1.40\tabularnewline
\hline 
$\Delta\phi_{o}$ (deg)&
$1.58\times10^{3}$&
$2.80\times10^{3}$&
33.3&
43.9\tabularnewline
\hline 
$\Delta\psi$ (deg)&
$1.62\times10^{3}$&
$2.87\times10^{3}$&
79.9&
111\tabularnewline
\hline 
$\Delta t_{c}$(sec)&
14.7&
77.2&
13.1&
68.5\tabularnewline
\hline
\end{tabular}\par\end{centering}
\end{table}

\section{Conclusion\label{sec:Conclusion}}

The results presented above indicate that inclusion of the merger
signal can have a significant impact on LISA's ability to measure
source parameters. It is interesting to compare this effect with other
effects that have been studied previously. In Table \ref{tab:LocCompare},
we compare our results for $M=10^{6}\,\mbox{M}_{\odot}$ at $z\approx1$
with prior published works. The parameters compared are the fractional
luminosity distance $\Delta D_{L}/D_{L}\approx\Delta\ln(D_{L})$ and
the solid angle subtended by the error ellipse, $\Delta\Omega_{N}$.
The latter quantity can be computed from the sky position angles,
their errors, and their covariance. Aside from the differences in
the waveform physics (spin, merger, etc.), the biggest difference
in the studies is the length of the waveforms. Ours are a factor of
$\sim30$ shorter than those used in the other published works due
to the computational effort required to simulate long waveforms with
\emph{Synthetic LISA}. 

The best comparison that can be made with this work is to compare
the non-spinning results of Cutler and of Vecchio to our pre-ISCO
case. $\Delta D_{L}/D_{L}$ is a factor of $2\sim5$ larger in our
results than the published results. When the merger is added, the
three results for $\Delta D_{L}/D_{L}$ approximately agree. This
suggests that observing the final $\sim10\,\mbox{days}$ of a $M=10^{6}\,\mbox{M}_{\odot}$
MBHB at $z\sim1$ will constrain $D_{L}$ equally as well as observing
the entire year prior to ISCO. 

The results for $\Delta\Omega_{N}$ for the short waveforms are considerably
worse. This is consistent with the picture that modulations due to
LISA's orbital motion provide much of the sensitivity to these parameters.
Nevertheless, the improvement in $\Delta\Omega_{N}$ upon inclusion
of the merger is a factor of $\sim5$, roughly the same factor as
the inclusion of spin precession. 

Whether this same level of improvement will persist when long-duration
observations, spin, and precession are added will depend on the exact
mechanism by which the improvement arises. If both improvements result
from breaking the same parameter degeneracy, then the improvement
is likely to be minimal. If however, the mechanisms for improvement
are orthogonal, it is possible that the relative improvement upon including
the merger will be similar to the case here. 

\begin{table}

\caption{\label{tab:LocCompare}Comparison of source location for equal-mass
MBHBs at $z\sim1$ in this work to prior published works. S,P, and
M refer to waveforms with spin, spin precession, and merger respectively.
($^{*}$ estimated from the results in Table II of \cite{Cutler_Resolution},
$^{\dagger}$ estimated from histograms in FIG. 2 of \cite{Vecchio_Spin})}

\begin{centering}\begin{tabular}{|c|c|c|c|c|c|c|c|}
\hline 
&
$M\,(10^{6}\,\mbox{M}_{\odot})$ &
duration (yr)&
S&
P&
M&
$\Delta\Omega_{N}$ ($\mbox{deg}^{2}$)&
$\Delta D_{L}/D_{L}$\tabularnewline
\hline
\hline 
Cutler\cite{Cutler_Resolution}&
4.0&
1.0&
no&
no&
no&
$4\times10^{-1*}$&
$7\times10^{-2*}$\tabularnewline
\hline 
Vecchio\cite{Vecchio_Spin}&
4.0&
1.0&
no&
no&
no&
$6\times10^{-1\dagger}$&
$2\times10^{-2\dagger}$\tabularnewline
\hline 
Vecchio\cite{Vecchio_Spin}&
4.0&
1.0&
yes&
no&
no&
$3\times10^{-1\dagger}$&
$6\times10^{-3\dagger}$\tabularnewline
\hline 
Lang \& Hughes\cite{LangHughes_Localizing}&
1.2&
$\sim$1.5 (avg)&
yes&
yes&
no&
$5.56\times10^{-2}$&
$3.57\times10^{-3}$\tabularnewline
\hline 
this work &
1.0&
0.03&
no&
no&
no&
$5.08\times10^{1}$&
$1.18\times10^{-1}$\tabularnewline
\hline 
this work&
1.0&
0.03&
no&
no&
yes&
$1.13\times10^{1}$&
$4.48\times10^{-2}$\tabularnewline
\hline
\end{tabular}\par\end{centering}
\end{table}

\section*{Acknowledgments}

Copyright (c) 2008 United States Government as represented by the
Administrator of the National Aeronautics and Space Administration.
No copyright is claimed in the United States under Title 17, U.S.
Code. All other rights reserved. This research was supported by an
appointment to the NASA Postdoctoral Program at the Goddard Space
Flight Center, administered by Oak Ridge Associated Universities through
a contract with NASA.


\section*{References}
\bibliographystyle{unsrt}
\bibliography{Thorpe_LISA7_final}
\end{document}